\documentclass[11pt]{article}
\usepackage{amssymb}
\usepackage{amsmath}
\usepackage{texdraw}

\newcommand{\dif}{\mbox{\rm{d}}}

\begin{document}

\title{\bf Generalized hyperbolic Ernst equations for an Einstein-Maxwell-Weyl field}
\author{Anastasios Tongas and Frank Nijhoff \\
{\em \normalsize Department of Applied Mathematics, University of Leeds, Leeds LS2 9JT, U.K.}}

\maketitle

\begin{abstract} 
A new family of nonlinear partial differential equations is presented.
They represent a generalization of the hyperbolic Ernst equations for an Einstein-
Mawxell-Weyl field in general relativity. A B\"acklund transformation 
for the system of equations under consideration is given, and
their direct relation to the complete Boussinesq hierarchy of soliton equations
is illustrated.
\end{abstract}

\section{Introduction}

Plane symmetric spacetimes are of fundamental importance in Einstein's theory
of gravitation. They describe the nonlinear interaction following the collision
of plane fronted gravitational waves solely, or coupled with other fields.
In this context the basis of the vacuum Einstein
equations is formed by the hyperbolic version of the famous Ernst equation
\cite{Ernst1,Ernst2}. Many exact solutions of the Ernst equation and of its various
extensions, which form the basis of the Einstein equations in the
presence of certain fields, have been found \cite{Griffithsbook}.
These solutions allow one to study the collisional interactions, indirectly.
Assuming that the approaching waves are known, the direct problem is formulated
by the characteristic initial value problem of the hyperbolic Ernst equations.
For recent progress on the analysis of the preceding problem, as well as on
the development of an effective integral representation of the solution in the
interaction region, we refer to \cite{fst,HE,Alex1,AlexGrif}. The underlying
structure which has lent to the Ernst equation a prominent
position as a nonlinear partial differential equation is the large number of
the internal symmetries, as was originally suggested by Geroch \cite{Geroch1,Geroch2}.
More recently in \cite{Schief} it was shown that the infinitesimal Geroch
transformations for stationary axisymmetric spacetimes extend to a large class
of equations which reside in the $(2+1)$-dimensional Loewner system. The latter
system encodes a variety of equations which from a modern perspective are
considered to be integrable.

One of the outstanding universal features of integrable nonlinear evolution
equations is the fact that they are not isolated objects, but they arise always
with a structured infinite dimensional hierarchy of mutually compatible evolution
equations. The latter has been exploited in many ways, leading to the intimate
connection between the associated hierarchy of an integrable evolution
equation and the representation theory of affine Lie algebras (and loop groups),
see e.g. \cite{Kac} and references therein. The point of view
that one single equation in the hierarchy possesses a dominant character is
misguided, and in fact, one should consider the infinite hierarchy of all
equations together as being the integrable system.

It is remarkable that there exist circumstances in which the entire hierarchy of
an integrable evolution equation can be encoded in a single
equation. A prime example of such an equation was given in \cite{NHJ}, and it
is represented by a scalar fourth order partial differential equation (PDE).
The entire hierarchy of the Korteweg-de Vries (KdV) equations follows by
systematic expansions from this scalar PDE. Moreover, in \cite{TTX} it was shown
that a proper generalization of this PDE incorporates the hyperbolic Ernst equation
for a Weyl neutrino field in general relativity.

In this paper, a novel system of PDEs of fourth order in the independent variables is presented, 
which incorporates the hyperbolic Ernst equations for a source-free Maxwell field and a Weyl neutrino field.
On the other hand this system represents in fact a {\em consistent 
parameter-family} of PDEs in the same spirit reported in \cite{NHJ}, 
which encodes the entire hierarchy of Boussinesq soliton equations. 
To be more precise, the system of PDEs arises as the Euler-Lagrange equations
for the variational problem associated with the Lagrangian density
\begin{equation}
L = (u-v)\,\frac{\mathcal{F} \mathcal{E}}{\mathcal{J}^2} + m(v)\, \frac{\mathcal{F}}{\mathcal{J}} + 
n(u)\, \frac{\mathcal{E}}{\mathcal{J}}\,.
\label{eq:Lagr}
\end{equation}
The scalars quantities $\mathcal{F},\mathcal{E}$ and $\mathcal{J}$ are given by
\begin{equation}
\mathcal{F}= {{\boldsymbol f}}_{\!\!,uv} \wedge {{\boldsymbol f}}_{\!\!,u}\,, \quad
\mathcal{E}= {{\boldsymbol f}}_{\!\!,uv} \wedge {{\boldsymbol f}}_{\!\!,v}\,, \quad
\mathcal{J}= {{\boldsymbol f}}_{\!\!,u} \wedge {{\boldsymbol f}}_{\!\!,v}\,,
\label{eq:FJE}
\end{equation}
where $\boldsymbol{f}=(f_1,f_2)$. The components of $\boldsymbol{f}$,
are complex functions of $(u,v)\in \mathbb{C}^2$, and $n$ and $m$ are complex
parameter functions of the indicated arguments, unless otherwise stated. Partial
differentiation is denoted by a comma or $\partial$, followed by the
variable(s) with respect to which the differentiation has been performed.
The Lagrangian $L$ remains invariant, molulo $L$ and null Lagrangians,
under the projective group of linear fractional transformations on
$\boldsymbol{f}$, i.e.
\begin{equation}
{f_i}\,\, \longmapsto\,\, \frac{\alpha_{i1} f_1 + \alpha_{i2} f_2 +
\alpha_{i3}}{\alpha_{31} f_1 + \alpha_{32} f_2 +\alpha_{33}}\,,
\end{equation}
$i=1,2,$ and $(\alpha_{ij})\in\mathrm{SL}(3,\mathbb{C})$. If in addition
the parameter functions $n$ and $m$ are constants, then $L$ remains invariant
under the following affine transformations:
\begin{equation}
(u,v) \mapsto (\epsilon_1 u + \epsilon_2,\,\,\epsilon_1 v + \epsilon_2)\,,
\,\qquad \epsilon_1,\epsilon_2 \in
\mathbb{C},\,\, \epsilon_1 \neq 0. \label{eq:affine}
\end{equation}
The above symmetries are called divergence symmetries of the associated
variational problem, and every symmetry of
this type is inherited as a Lie-point symmetry to the Euler-Lagrange equations,
cf \cite{Olver}.

The integrability of the Euler-Lagrange equations associated with $L$, will be
demonstrated by the existence of the usual paraphernalia of the integrable models,
such as of an associated linear representation (Lax pair) and a B\"acklund
transformation, in sections 2 and 4, respectively. The connection with the complete
hierarchy of the Boussinesq soliton equations is addressed in section 3, by exploiting
the infinite symmetries of the field equations associated with the Lagrangian $L$.
Section 5 deals with the reduction of the field equations to the Ernst equations of
general relativity and the paper concludes in section 6 with discussion. 

The continuous field equations and the derivation of the associated Lax pair 
and B\"acklund transformation originate in a compatible parameter family, 
of multi-field partial difference equations (P$\Delta$Es); the two-dimensional 
discrete Boussinesq system. For the relevant constructions we use an intrinsic 
property which enjoys the discrete Boussinesq system and consists of its
{\em multi-dimensional consistency}. Recently in \cite{Adler}, this property has
been successfully exploited, to obtain a classification of one-field equations
living on elementary quadrilateral faces. It can be argued that discrete integrable 
systems are equally important just as their continuous analogues, and in many 
respects their study has led to new insights into the structures behind 
the more familiar continuous systems. The notion of B\"acklund 
transformations and their associated Bianchi permutability theorems, which emerged in
the transformation theory of surfaces in ordinary space in the late nineteenth century,
provide an excellent instance of such a consistency.
Thus, the investigations of the interplay between the discrete 
and the continuous integrable systems acquire a certain significance, and in our
view, the role of the second example of a generating PDE, becomes key.

\section{The field equations and their linear representation}

We first note that the Euler-Lagrange equations associated with the Lagrangian
$L$, admit of a linear representation.
This means that the field equations are a compatibility condition for the
existence of the solution of an auxiliary, linear, overdetermined system of
equations of the form
\begin{equation}
\psi_{,u} = A(u,v;\lambda) \psi\,, \quad \psi_{,v} = B(u,v;\lambda) \psi \,.
\label{eq:LA}
\end{equation}
The complex $r$-vector $\psi$ and the $r\times r$-matrices $A$, $B$ are functions
of $u$, $v$ and a complex parameter $\lambda$, called the spectral parameter.
The compatibility condition of the
overdetermined linear system (\ref{eq:LA}) is
\begin{equation}
A_{,v}-B_{,u} + \left[A,B\right]=0\,, \label{eq:zerocurv}
\end{equation}
which should hold identically for every value of $\lambda$, where $[\;\,,\,]$ denotes
usual matrix commutator. Let us now consider
the case where $A$ and $B$ are rank 1 matrices and their sole dependence
on the spectral parameter $\lambda$ is given by the following rational
expressions:
\begin{equation}
A_{ij}=\frac{a_i\, d_j}{u-\lambda},\qquad B_{ij} =\frac{b_i\, c_j}{v-\lambda}\,,
\label{eq:AB}
\end{equation}
$i,j=0,1\ldots r-1$. Then, equation (\ref{eq:zerocurv}) implies the existence of
the parameter functions $n(u)$, $m(v)$, introduced by
\begin{equation}
n(u)=a^i \, d_{i}\,,\quad m(v)=b^i \, c_{i}\, , \label{eq:nm1}
\end{equation}
$i=0,1,\ldots r-1$. Indices will be raised and lowered using the Kronecker delta
symbol $\delta_{ij}$, and summation over repeated upper and lower indices is
understood. Without loosing generality we may fix $a_0=b_0=1$. Consequently,
equation (\ref{eq:zerocurv}) leads to the following system of PDEs:  
\begin{eqnarray}
(u-v)\,a_{i,v} - a^j\, c_j \, (a_i-b_i) &=&0,\label{eq:PDE1}\\
(u-v)\,b_{i,u} - b^j\, c_j\,(a_i-b_i) &=&0,\label{eq:PDE2}\\
(u-v)\,d_{i,v} + a^j\, c_j\, d_i - b^j d_j c_i&=&0\,, \label{eq:PDE3}\, \\
(u-v)\,c_{i,u} + a^j\,c_j\, d_i - b^j d_j\, c_i&=&0 \,,  \label{eq:PDE4} 
\end{eqnarray}  
$i,j=0,1,\ldots r-1$. An obvious consequence of the latter system of PDEs
is the relation $c_{i,u}=d_{i,v}$, which implies the existence of the functions
$f_i$, such that
\begin{equation}
c_i = f_{i,v}\quad d_i = f_{i,u}\,, \label{eq:cd}
\end{equation}
$i=0,1\ldots r-1$. In terms of the new variables $f_i$ and using equation
(\ref{eq:nm1}), the system of PDEs (\ref{eq:PDE1})-(\ref{eq:PDE4}) reads  
\begin{eqnarray} \label{vfe1}
(u-v) a_{i,v} - \big(m + (a_j-b_j)\,\, f^j_{,v}\big) (a_i-b_i) &=&0,
\label{eq:sys1}\\
(u-v) b_{i,u} - \big(n - (a_j-b_j)\,\, f^j_{,u}\big) (a_i-b_i) &=&0,
\label{eq:sys2}\\
(u-v) f_{i,uv} + m f_{i,u} - n f_{i,v} + C_{ij} (a^j-b^j) &=&0\,,
\label{eq:sys3}
\end{eqnarray}  
where $i,j=1,2,\ldots r-1$, and
\begin{equation}
C_{ij}=f_{i,u}\,f_{j,v} + f_{i,v}\,f_{j,u}.
\end{equation}
We note that in general, the $(r-1)\times (r-1)$-matrix $C$ has rank $2$.
In the following we consider the case $r=3$. In this case, by eliminating
the variables $a_i$, $b_i$ from the system (\ref{eq:sys1})-(\ref{eq:sys3}),
we arrive at a coupled system of
fourth order PDEs for the variables $f_i$, $i=1,2$. To this end, one solves the
linear system (\ref{eq:sys3}) for the differences $a_i-b_i\equiv z_i$,
in terms of the partial derivatives of the unknown functions $f_i$.
Substituting the result into the remaining equations (\ref{eq:sys1}),
(\ref{eq:sys2}) and using the compatibility condition $b_{i,uv}=b_{i,vu}$
(or equivalently $a_{i,uv}=a_{i,vu}$) we get
\begin{equation}
D_u D_{v}\, z_i =
D_u \left(\frac{z_i (z^j f_{j,v}+m)}{u-v}\right) +
D_v \left(\frac{z_i (z^j f_{j,u}-n)}{u-v}\right)  \label{eq:ELeqs}
\end{equation}
$i,j=1,2$, where $D$ denotes total differentiation with respect to the
indicated arguments in the subscripts. The fourth order system of PDEs
(\ref{eq:ELeqs}), is precisely the equations obtained by varying the fields
$(f_1,f_2)$ in the Lagrangian $L$.

\section{Infinite symmetries and the Boussinesq hierarchy}

We next illustrate the infinite dimensional flows which are compatible with
the linear representation (\ref{eq:LA}), and consequently with the field
equations. This amounts to finding and representing the infinite
symmetries of the field equations. To this end we follow the construction
given in \cite{N1}, cf \cite{N2} in connection with techniques such as the
vertex operator. Having found the infinite symmetries of the field equations,
the representation is accomplished in three steps. First, by introducing an infinite 
set of indeterminates $x^i=(x^1,x^2,\ldots)$, which are the parameters of
the group associated with the symmetry algebra. Next, by imposing certain 
linear differential relations for the potentials $\psi_i$ with respect to 
$x^i$, in such a way that each of these relations is compatible with the linear pair
(\ref{eq:LA}). Finally, by treating the indeterminates $x^i$ as additional
independent variables, which leads naturally our considerations to the commuting
compatible flows, i.e. an infinite compatible set of PDEs for $H(u,v,x^i)$.

In order to find the symmetries of the field equations
(\ref{eq:sys1})-(\ref{eq:sys3}), it is necessary to introduce
the matrix-valued function $H$, as follows
\begin{equation}
{H}_{ij,u} = a_i \, f_{j,u}\,,\quad {H}_{ij,v} = b_i \, f_{j,v} \,,
\label{eq:Hpotential}
\end{equation}
$i,j=0,1,2$. The existence of $H$ is guaranteed by the field equations
(\ref{eq:sys1})-(\ref{eq:sys3}), which now acquire the matrix form
\begin{equation}
(v-u)\,{H}_{,uv} + [H_{,u},H_{,v}] = 0. \label{eq:H}
\end{equation}
The symmetry algebra of equation (\ref{eq:H}) is generated by
all characteristic vector fields
\begin{equation}
X_Q = Q\, \partial_H + (D_u Q) \,\partial_{H_{,u}} + (D_v Q) \,\partial_{H_{,v}}
+(D_u D_v Q) \,\partial_{H_{,uv}}\,,
\end{equation}
for which the matrix-valued function $Q$ satisfies the linearized
equation (\ref{eq:H}), namely
\begin{equation}
(v-u)\,D_u D_v Q + [D_u Q, H_{,v}] + [H_{,u}, D_v Q] = 0\,. \label{eq:det}
\end{equation}
Equation (\ref{eq:det}) should hold for all solutions of equation (\ref{eq:H}), and
as such the field equations and their differential consequences
are taken into account. In general, the symmetry characteristic $Q$ depends on
the coordinates $(u,v)$, on $H$ and its partial derivatives up to an unspecified
order, and on potentials of $H$.
If $Q$ is linear in the first partial derivatives of $H$, then equation
(\ref{eq:det}) delivers the local, or Lie-point, symmetries of equations (\ref{eq:H}).
These symmetries are given by the characteristic
\begin{equation}
Q_0=\varepsilon_1 \left(\,[J,H]+K\,\right)
+ \varepsilon_2 \left(\, H_{,u} + H_{,v}\,\right) +
\varepsilon_3 \left(\, u H_{,u} + v H_{,v}\,\right)\,,
\end{equation}
where $J$ and $K$ are complex constant matrices and $\varepsilon_i$ complex constants.
It should be noticed that if $n$ and $m$ are parameter functions of $u$ and $v$ respectively,
then only the symmetry with $\varepsilon_1 \neq 0$ and $\varepsilon_2=\varepsilon_3=0$,
is compatible with the symmetries of the field equations (\ref{eq:sys1})-(\ref{eq:sys3}),
since on the contrary the characteristic corresponds to the affine base transformations
(\ref{eq:affine}). The following observation allows one to find in principle, an infinite
number of symmetries of equations (\ref{eq:H}), starting from the characteristic
$Q_0$ with $\varepsilon_1=1,\varepsilon_2=\varepsilon_3=0$.

Equation (\ref{eq:det}), can be written in the form
\begin{equation}
D_v \left(u D_u Q + [Q, H_{,u}]\right) =
D_u \left(v D_v Q + [Q, H_{,v}]\right)\,,
\label{eq:detcons}
\end{equation}
which implies the existence of the matrix-valued
potential $\widetilde{Q}$ introduced by
\begin{equation}
D_u \widetilde{Q} =  u D_u Q + [Q, H_{,u}]\,, \qquad
D_v \widetilde{Q} =  v D_v Q + [Q, H_{,v}].
\label{eq:Rpotential}
\end{equation}
By direct calculations and using the Jacobi identity, one finds that the
potential $\widetilde{Q}$ is also a symmetry
of equations (\ref{eq:H}), whenever $Q$ and $H$ satisfy (\ref{eq:det}) and
(\ref{eq:H}), respectively. In other words,
equations (\ref{eq:Rpotential}) define a recursion operator $\mathcal{R}$,
i.e. an operator with the property that whenever $Q$ is a characteristic
of a symmetry of equations (\ref{eq:H}), so is $\widetilde{Q}=\mathcal{R}Q$.
It follows from equations (\ref{eq:Rpotential}), that we may formally write
$\mathcal{R}$ as the operator
\begin{equation}
\mathcal{R}= {\mathsf{D}}^{-1}\cdot (\, \sigma\, {\mathsf{D}} -
\varrho \ast {\mathsf{D}} + \mathrm{ad}_{\mathrm{d}H}) \,.
\end{equation}
Here, $\mathsf{D}$ and $\mathrm{d}$ denote total and ordinary differentials respectively,
$\ast$ is the two dimensional Hodge duality operator
which acts on the basis of one forms by
\begin{equation}
\ast \dif u = \dif u,\quad \ast \dif v = -\dif v\,,
\end{equation}
$\mathrm{ad}_{A}(B)=[B,A]$, and the functions $\varrho$ and
$\sigma$ are given by
\begin{equation}
\sigma=\textstyle{\frac{1}{2}(v+u)\,, \quad \varrho=\frac{1}{2}(v-u)\,.}
\end{equation}

From the above discussion it is clear that starting with specific matrices
$J$ and $K$ and using the recursion relations
\begin{equation}
Q_{p+1}={\mathcal{R}}Q_p, \quad Q_0=[J,H] + K,\quad p \in \mathbb{N}\,,
\end{equation}
we may generate an infinite number of symmetries of equations (\ref{eq:H}).
Let $x^p=(x^0,x^1,\ldots)$, $p\in \mathbb{N}$, denote the parameters of the
infinitesimal symmetry transformations which are generated by the 
vector fields $X_{Q_p}$, i.e.
\begin{equation}
\partial_{x^p} H = Q_{p}, \quad p \in \mathbb{N}\,. \label{eq:group}
\end{equation}
Since for every fixed $p$, the characteristic $Q_p$ defines a symmetry of equation (\ref{eq:H}), 
the pair of PDEs which is formed by taking each of equations (\ref{eq:group}) together 
with equation (\ref{eq:H}), is a compatible system of PDEs for $H(u,v,x^p)$ ($p$ fixed). 
By virtue of this fact, every such pair of PDEs, admits also a linear representation. 
The latter is given by the linear equations (\ref{eq:LA}) for $\psi$, and which for 
convenience we rewrite below
\begin{equation}
(u-\lambda) \psi_{,u} = H_{,u}\, \psi\,, \qquad
(v-\lambda) \psi_{,v} = H_{,v}\, \psi\,, \label{eq:LAH}
\end{equation}
together with each of the following linear equations 
\begin{eqnarray}
\psi_{,x^0} &=& J\,\psi\,,  \label{eq:hp1}\\
\psi_{,x^{p+1}} &=& -\lambda\, \psi_{,x^{p}} - Q_{p}\, \psi\,, \qquad p \in \mathbb{N}\,.
\label{eq:hp2}
\end{eqnarray} 
By induction, we may easily prove that the compatibility conditions 
$\partial_{x^p} \partial_u \psi = \partial_u \partial_{x^p} \psi$ and
$\partial_{x^p} \partial_v \psi = \partial_v \partial_{x^p} \psi$ are satisfied 
for every $p\in \mathbb{N}$, whenever $H$ satisfies equation
(\ref{eq:H}) and $Q_p$ are the symmetry characteristics of the previous construction.

Similarly, starting with some other matrices $\widehat{J}$ and $\widehat{K}$, one obtains 
a different infinite hierarchy of symmeries of equation (\ref{eq:H}). Let 
$y^q=(y^0,y^1,\ldots)$, $q\in \mathbb{N}$ denote the parameters of the 
corresponding symmetry transformations generated by the vector fields 
$X_{\widehat{Q}_{q}}$. In analogy with the preceding construction, the linear equations 
which can be imposed in addition to the linear equations (\ref{eq:LAH}) are  
\begin{eqnarray}
\psi_{,y^0} &=& \widehat{J}\,\psi\,,  \label{eq:hq1}\\
\psi_{,y^{q+1}} &=& -\lambda\, \psi_{,y^{q}} - \widehat{Q}_{q}\, \psi\,, \qquad q \in \mathbb{N}\,,
\label{eq:hq2}
\end{eqnarray} 
We now assume that the potentials $H$ and $\psi$ depend in addition to $(u,v)$, 
on $x^p$ and also on $y^q$, $\forall p,q\in \mathbb{N}$. Since $x^p$ and
$y^q$ will be treated as additional independent variables, it is necessary that
\begin{equation}
\partial_{x^p} \partial_{y^q} F -  \partial_{y^q} \partial_{x^p} F = 0\,,
\qquad \forall \, p,q\in \mathbb{N}\,.
\label{eq:comhier}
\end{equation} 
holds, for all functions $F(u,v,x^p,y^q)$.
The aim now is to {\em generate} the compatible set of PDEs 
for $H$, in terms of the higher {\em time variables} $x^p$, $y^q$. This infinite set is
obtained by taking the compatibility conditions 
$\partial_{x^p} \partial_{y^q} \psi = \partial_{y^q} \partial_{x^p} \psi$, on the 
linear representations (\ref{eq:hp1}), (\ref{eq:hp2}) and (\ref{eq:hq1}), (\ref{eq:hq2}).
In order to give an explicit description and for purposes of calculation we
proceed as follows.
  
The leading equation (\ref{eq:comhier}) is valid for every solution $H$ of equation (\ref{eq:H}), 
whenever the starting matrices satisfy 
\begin{equation}
[J,\widehat{J}]=0\,,\,\quad \mbox{and}\quad [J,\widehat{K}] = [\widehat{J},K]\,. \label{eq:com}
\end{equation}
By virtue of relations (\ref{eq:com}), we fix the values of the starting matrices as follows:
\begin{eqnarray}
(J_{ij})= \left(\begin{array}{rcc}
0 & 0 & 0 \\
0 & 0 & 0 \\
1 & 0 & 0 \end{array}\right) \,, \quad
(K_{ij})= \left(\begin{array}{rrr}
0 & -1 & 0 \\
0 & 0 & -1 \\
0 & 0 & 0 \end{array}\right) \,, &
\\
(\widehat{J}_{ij}) = \left(\begin{array}{rcc}
0 & 0 & 0 \\
1 & 0 & 0 \\
0 & 1 & 0 \end{array}\right) \,, \quad
(\widehat{K}_{ij})= \left(\begin{array}{rrr}
0 & \phantom{-}0 & -1 \\
0 & 0 & 0 \\
0 & 0 & 0 \end{array}\right) \,,
\end{eqnarray} 
$i,j=0,1,2$. Next, we factor out the linear set of PDEs for $H$, which is obtained from
the compatibility conditions $\partial_{x^p} \partial_{y^q} \psi = \partial_{y^q} \partial_{x^p} \psi$
with $p=0,q\geq 1$, ($q=0,p\geq 1 )$. Finally, we
combine the remaining linear representations (\ref{eq:hp2}) and (\ref{eq:hq2})
into one, by introducing (i) auxiliary higher time variables 
\begin{equation}
w^i=(w^1=x^1,w^2=y^1,w^4,w^5,w^7\ldots),\quad 
i \in {\mathbb Z}\setminus 3{\mathbb Z},\,\,\, i\geq 1,
\end{equation}
and (ii) the matrix 
\begin{equation}
\Lambda=-\lambda J - K\,.
\end{equation} 
From this starting point, and noticing that 
\begin{equation}
\Lambda^2=-\lambda \widehat{J} - \widehat{K}\,,
\end{equation}
the linear infinite set of differential relations for $\psi$, which can be
{\em simultaneously} imposed on the linear representation (\ref{eq:LAH}), is
given by
\begin{eqnarray}
\psi_{,w^1} &=& (\Lambda\,\,\, +\, [\,\,\Lambda^{\prime}\,, {H}\,]\,)\psi\,,
\label{eq:hierarchy1}\\
\psi_{,w^2} &=& (\Lambda^2 +\, [\,{(\Lambda^2)}^{\prime}, {H}\,]\,)\psi\,,
\label{eq:hierarchy2} \\
\psi_{,w^{i+3}} &=& -\lambda\, \psi_{,w^i}\, - \, {H}_{,w^i}\,\psi\,, 
\qquad i \in {\mathbb Z} \setminus 3 {\mathbb Z}, \quad i\geq 1\,,
\label{eq:hierarchyj}
\end{eqnarray} 
where prime in equations (\ref{eq:hierarchy1}), (\ref{eq:hierarchy2}) denotes differentiation with 
respect to $\lambda$. The above differential relations generate an infinite 
set of PDEs for $H$, with the Boussinesq equation being the first 
member. Specifically, the compatibility condition $\partial_{w^2} \partial_{w^1} \psi = 
\partial_{w^1} \partial_{w^2} \psi$  
on equations (\ref{eq:hierarchy1}), (\ref{eq:hierarchy2}), yields a set of six, first order PDEs for 
the components of $H$. After a simple differentiation and elimination process 
applied to the latter system, and using (\ref{eq:Hpotential}) to identify the 
main field variables $(f_1,f_2)$, we arrive at the following system of PDEs:    
\begin{eqnarray} 
f_{1,xy} &=& \textstyle{2 f_{2,yy} + (f_2 f_{2,x})_{,y} - 
6 f_{2,x} f_{2,xx} - \frac{1}{2}f_{2,xxy} + 
\frac{1}{4}f_{2,xxxx} }\,, \label{eq:fxy} \\ 
f_{1,xx} &=& \textstyle{(f_2 f_{2,x})_{,x} + \frac{1}{2}(f_{2,xy} - 
f_{2,xxx})}\,, \label{eq:fxx}
\end{eqnarray} 
where $x=w^1,y=w^2$. Finally, taking the compatibility condition 
$f_{1,xyx}=f_{1,xxy}$ on equations (\ref{eq:fxy}), (\ref{eq:fxx}) we obtain
\begin{equation}
3 {\mathcal{B}}_{,yy}+ {\mathcal{B}}_{,xxxx} + 6 ({\mathcal{B}}^2)_{,xx} = 0 \, , \label{eq:BSQ}
\end{equation}
where ${\mathcal{B}}=-f_{2,x}$. Equation (\ref{eq:BSQ}) is known as the Boussinesq 
equation. Similarly, higher members of the 
Boussinesq hierarchy can be derived from the system 
(\ref{eq:hierarchy1})-(\ref{eq:hierarchyj}), in a systematic way.

\section{A B\"acklund auto-transformation}
The discussion in the present section relates to matters that had their origin in 
the investigations by B\"acklund near the end of the nineteenth century, concerning the 
transformation properties of simultaneous partial differential equations and the 
transformation theory of surfaces in ordinary space.
There are many methods developed over the past century to obtain
the B\"acklund transformations for certain classes of partial 
differential equations, see \cite{AI,RS1,Mat,RS2} for recent advances on these topics.
However, for the construction of a B\"acklund auto-transformation 
for the relevant PDEs of our discussion, i.e. a transformation within the space of solutions of
PDEs (\ref{eq:sys1})-(\ref{eq:sys3}), we find the connection with the lattice
Boussinesq equation, as a consequence of its three dimensional consistency, as
the most powerfull.

An upshot of recent investigation of the equations (\ref{eq:ELeqs}), and their 
compatible discrete counterparts, has been the insight that both systems arise 
as parameter families of {\em mutually} consistent equations. In turn, these 
equations form multi-variable and multi-parameter systems, in which parameters 
and variables play a dual (even interchangeable) role, and the whole infinite 
system is characterized by an intrinsic consistency. 
In the light of the close relation of equation (\ref{eq:ELeqs}) with the 
Ernst equations, as will be explained in the next section, and since solution 
generating techniques such as B\"acklund transformations, have proven to be 
invaluable methods in finding physically significant exact solutions of 
the Ernst equations \cite{Harrison,Neu,KramerNeu}, we now explain how a B\"acklund 
auto-transformation for equations (\ref{eq:ELeqs}) can be constructed straightforward,
by exploiting the consistency approach. 

The construction is based on the compatible discrete system as this 
was derived in \cite{TN}. In order to make contact with the latter system, it is necessary 
to retract for the moment the dependence of the parameter functions $n$, $m$ on $u,v$, respectively.  
\footnote{Relative to the continuous field equations, this restriction is not essential
since the B\"acklund transformation that we derive below, is also valid with $n,m$ parameter functions.} 
Let us now consider an elementary {\em plaquette} of a two-dimensional lattice, as shown 
in figure 1. The dependent variables (fields) are assigned on the vertices at sites 
$(n,m)$, and $(u,v)$ which now play the role of the lattice parameters, are assigned on the edges.
The updates of a lattice variable $F$ along a shift in the $n$ and 
$m$ direction of the lattice will be denoted by $F_{[1]},F_{[2]}$ respectively, i.e.
\begin{equation}
F_{[1]}=F(n+1,m),\quad F_{[2]}=F(n,m+1),\quad F_{[12]}=F(n+1,m+1)\,. \label{eq:shifts}
\end{equation}
\begin{figure}[h]
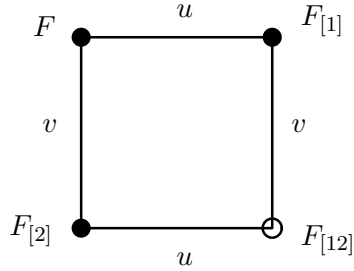

\centertexdraw{
\setunitscale 0.5
\linewd 0.03 \arrowheadtype t:F 
\move (-1 -1)
\lvec (-1 1) \lvec (1 1) \lvec (1. -1) \lvec(-1 -1)
\move(-1 -1) \fcir f:0.0 r:0.1 \move(-1 1) \fcir f:0.0 r:0.1 
\move(1 1) \fcir f:0.0 r:0.1 \move(1 -1) \lcir  r:0.1    

\htext (-1.75 -1.2) {$F_{[2]}$}  
\htext (-1.5 1) {$F$}
\htext (1.3 1) {$F_{[1]}$} 
\htext (1.3 -1.3) {$F_{[12]}$} 
\htext (0 1.2) {$u$} 
\htext (0 -1.4) {$u$} 
\htext (-1.4 0) {$v$} 
\htext (1.2 0) {$v$} 
}
\caption{An elementary quadrilateral for the discrete Boussinesq system.}
\label{fig:BSQ}
\end{figure}
The partial difference equations which are compatible with the partial differential equations 
(\ref{eq:sys1})-(\ref{eq:sys3}), 
in the sense that these two systems share a common set of solutions, is given by the following
parameter family of algebraic equations: 
\begin{eqnarray} 
W_{[1]} &=& U\, U_{[1]} - V \,, \label{eq:latticeBSQ1} \\  
W_{[2]} &=& U\, U_{[2]} - V\,, \label{eq:latticeBSQ2} \\
W &=& U\, U_{[12]} - V_{[12]} + \frac{u-v}{U_{[2]} - U_{[1]}}\,. 
\label{eq:latticeBSQ12} 
\end{eqnarray}
The system of equations (\ref{eq:latticeBSQ1})-(\ref{eq:latticeBSQ12}), relates the values of the fields
assigned on the four vertices of the quadrilateral, where now $F=(U,V,W)$, 
and is called the discrete Boussinesq system. 
For the Cauchy problem on a staircase, we should impose on the black points 
of the elementary {\em plaquette} of figure 1, the initial values
$(U,V,W)$, $(U_{[1]},V_{[1]})$, $(U_{[2]},V_{[2]})$ 
only, since from these data and the discrete equations (\ref{eq:latticeBSQ1})-(\ref{eq:latticeBSQ12}),
the values $(W_{[1]},W_{[2]})$ and $(U_{[12]},V_{[12]},W_{[12]})$ are determined uniquely.
In particular, using equations 
(\ref{eq:latticeBSQ1}), (\ref{eq:latticeBSQ2}) we find that
\begin{equation}
U_{[12]} = \frac{V_{[1]}-V_{[2]}}{U_{[1]}-U_{[2]}}\,,  \label{eq:latticeBSQ12U}
\end{equation}
and subsequently the values $W_{[12]}$ and $V_{[12]}$ can be found uniquely, from
equation (\ref{eq:latticeBSQ1}) (or (\ref{eq:latticeBSQ2})) and equation (\ref{eq:latticeBSQ12}), 
respectively. The fields which are assigned on the vertices with black points, are identified 
with the fields of the continuous PDEs, as follows:
\begin{eqnarray} 
W&=&f_1\,,\quad U\;\;=-f_2\,,\quad V\;=-H_{12}\,, \label{eq:id1}\\
U_{[1]}&=&a_1,\quad U_{[2]}=b_1 \qquad V_{[1]}=a_2,\qquad V_{[2]}=b_2\,. \label{eq:id2}
\end{eqnarray} 
In \cite{TN} it is shown that the discrete Boussinesq system
represents a consistent parameter family of two-dimensional lattice equations, in the sense
that equations (\ref{eq:latticeBSQ1})-(\ref{eq:latticeBSQ12}) can be imposed on all faces of a 
three-dimensional lattice in a consistent way. 
The third direction of the generated lattice has a dual role.
On the discrete level, is taken as auxiliary (spectral) and the system
is linearized in the corresponding variables $U_{[3]},V_{[3]}$ and their shifts.  
On the other hand, the discrete equations which relate the fields and their  
shifts in the third direction, define a map in the solution space 
of the continuous field equations, by virtue of the identifications (\ref{eq:id1})-(\ref{eq:id2}). 

Specifically, extending equations (\ref{eq:latticeBSQ1})-(\ref{eq:latticeBSQ12}) into a third 
direction associated with a new lattice parameter (identified here with the spectral 
parameter $\lambda$) we get
\begin{eqnarray}
W_{[3]} &=& U \, U_{[3]} - V \,,\\ \label{eq:latticeBSQ13}
U_{[13]} &=& \frac{V_{[1]}-V_{[3]}}{U_{[1]}-U_{[3]}}\,, \qquad
W = U\, U_{[13]} - V_{[13]} + \frac{u-\lambda}{U_{[3]} - U_{[1]}}\,, \label{eq:latticeBSQ23} \\
U_{[23]} &=& \frac{V_{[2]}-V_{[3]}}{U_{[2]}-U_{[3]}}\,, \qquad
W = U\, U_{[23]} - V_{[23]} + \frac{v-\lambda}{U_{[3]} - U_{[2]}}\,. \label{eq:latticeBSQ33}
\end{eqnarray} 
The auxiliary variables $U_{[3]},V_{[3]}$ are identified as projective variables and their relation
with the spectral variables $\psi_0,\psi_1,\psi_2$ is given by
\begin{equation}
U_{[3]}=\frac{\psi_1}{\psi_0}\,,\quad V_{[3]}=\frac{\psi_2}{\psi_0}\,. \label{eq:project}
\end{equation}
By straightforward calculations one easily verifies that the linear system 
(\ref{eq:LA}) for the potentials $(\psi_0,\psi_1,\psi_2)$,
implies the Riccati system for the potentials $(q_1,q_2)\equiv(U_{[3]},V_{[3]})$,
\begin{eqnarray}
(u-\lambda)q_{i,u}&=&(a_i-q_i)\big(n-(a_j-q_j)f^j_{,u}\big)\,, 
\label{eq:riccati1}\\
(v-\lambda)q_{i,v}&=&(b_i-q_i)\big(m-(b_j-q_j)f^j_{,v}\big)\,, 
\label{eq:riccati2}
\end{eqnarray} 
$i=1,2$. In virtue of the identifications (\ref{eq:id1})-(\ref{eq:id2}), 
we now interpret the shift of the fields in the third direction,
as a map $\varphi$ in the solution space of equations (\ref{eq:sys1})-(\ref{eq:sys3}). 
Explicitly, using equations (\ref{eq:latticeBSQ13})-(\ref{eq:latticeBSQ33}) and the identifications 
(\ref{eq:id1})-(\ref{eq:id2}), the map $\varphi$ is defined by   
\begin{eqnarray}
\varphi(f_1) = H_{12} -f_2\, q_1 \,, & &  \varphi(f_2) = -q_1 \,, \label{eq:BT1}\\
\varphi(a_1) = \frac{a_2-q_2}{a_1-q_1}\,,\quad\,\,\, & &
\varphi(a_2) =  - f_1  - f_2 \, \varphi(a_1) + \frac{u-\lambda}{q_1 - a_1}\,, \label{eq:BT2} \\
\varphi(b_1) = \frac{b_2-q_2}{b_1-q_1}\,, \quad\,\,\, & &
\varphi(b_2) =  - f_1  - f_2 \, \varphi(b_1) + \frac{v-\lambda}{q_1 - b_1}\,.  \label{eq:BT3}
\end{eqnarray} 
Restricting now our considerations on the continuum level, the above equations define a B\"acklund 
auto-transformation, i.e. a transformation that generates new solutions of the field equations 
from other known ones. Below we summarize the resulting solution generating procedure, 
which can be verified by lenghy but straightforward calculations.

Let $(f_1,f_2)$ be a known solution of equations (\ref{eq:ELeqs}). From this 
solution and using equations (\ref{eq:sys1})-(\ref{eq:sys3}), one determines by 
quadrature the corresponding vector components $(a_1,a_2)$ and $(b_1,b_2)$, up 
to constants of integration. Successively, the auxiliary complex potentials 
$H_{12}$ and $(q_1,q_2)$ can be found by quadrature from equation 
(\ref{eq:Hpotential}), and by integrating the Riccati system 
(\ref{eq:riccati1})-(\ref{eq:riccati2}), respectively.
Then, a new solution $({\widetilde a}_1,{\widetilde a}_2)$, 
$({\widetilde b}_1,{\widetilde b}_2)$ and $({\widetilde f}_1,{\widetilde f}_2)$
of the system (\ref{eq:sys1})-(\ref{eq:sys3}), and a fortiori for the equations 
(\ref{eq:ELeqs}), is given by the algebraic relations (\ref{eq:BT1})-(\ref{eq:BT3}), 
where ${\widetilde f}_i=\varphi(f_i)$, ${\widetilde a}_i=\varphi(a_i)$, ${\widetilde b}_i=\varphi(b_i)$,
$i=1,2$.

\section{Reduction to the Ernst equations}
We next show that the main second order subsystem of PDEs which is 
incorporated into the fourth order system of 
equations (\ref{eq:ELeqs}), is the hyperbolic Ernst equations for an 
Einstein-Maxwell-Weyl field. In order to make this connection more 
transparent, we introduce auxiliary complex potentials $(h_1,h_2)$ by
the relations
\begin{equation}
h_1= \frac{a_2-b_2}{a_1-b_1}\,,\qquad
h_2= \frac{b_1\,a_2-a_1\,b_2}{a_1-b_1}\,. \label{eq:h1h2}
\end{equation}
and which correspond to a change of the variables $(a_2,b_2)$, of the form
\begin{equation}
a_2= h_1\,a_1 + h_2\,,\qquad b_2= h_1\,b_1 + h_2\,. \label{eq:change}
\end{equation}
It follows from the system of equations (\ref{eq:sys1}), (\ref{eq:sys2}) that 
\begin{equation}
a_{2,v} = h_1\,a_{1,v} ,,\qquad b_{2,u} = h_1\,b_{1,u}\,, 
\end{equation}
which in virtue of equations (\ref{eq:change}), yield
\begin{equation}
h_{2,v} = -h_{1,v}\,a_1 \,, \qquad h_{2,u} = -h_{1,u}\, b_1\,. \label{eq:auxh2}
\end{equation}
From this starting point, and eliminating the variable $f_1$ from the field equations 
(\ref{eq:sys1})-(\ref{eq:sys3}), we may recast the latter system as a coulped set of 
four PDEs of second order in terms of
the variables $(a_1,b_1,h_1,f_2)$. This system of PDEs is given by 
\begin{eqnarray}
a_{1,uv} &=& 2\frac{a_{1,u} a_{1,v}}{a_1-b_1} - \frac{m}{u-v}\,a_{1,u}-\frac{n+1}{u-v}\,a_{1,v} + 
\frac{h_{1,u} f_{2,v}}{u-v} (a_1-b_1)^2\,,\label{eq:aux1}\\
b_{1,uv} &=& 2\frac{b_{1,u} b_{1,v}}{b_1-a_1} + \frac{m+1}{u-v}\,b_{1,u}+\frac{n}{u-v}\,b_{1,v} - 
\frac{h_{1,v} f_{2,u}}{u-v} (a_1-b_1)^2\,,\label{eq:aux2}\\
f_{2,uv}&=& \frac{1}{a_1-b_1} (b_{1,u} f_{2,v}- a_{1,v} f_{2,u})\,,\label{eq:aux3}\\
h_{1,uv}&=& \frac{1}{a_1-b_1} (b_{1,v} h_{1,u}- a_{1,u} h_{1,v})\,, \label{eq:aux4}
\end{eqnarray} 
where the first three PDEs follow from the system (\ref{eq:sys1})-(\ref{eq:sys3}), and
equation (\ref{eq:aux4}) follows from the compatibility condition $h_{2,uv}=h_{2,vu}$ on equations
(\ref{eq:auxh2}).

Let now $u,v$ be real valued, and $\star$ denote complex conjugation.
It follows from the system of PDEs (\ref{eq:aux1})-(\ref{eq:aux4}), that the 
latter admits a compatible reduction by applying the reality conditions
\begin{equation}
a_1 = - b_1^{\star}\,,\qquad h_1 = f_2^{\star} \, , \label{eq:reality}
\end{equation}
whenever the parameter functions $n$, $m$ satisfy
\begin{equation}
n + n^\star + 1=0\,,\quad m + m^\star + 1=0\, . \label{eq:nm}
\end{equation}
The resulting set of equations (\ref{eq:aux1}) and (\ref{eq:aux3}), which is formed
by applying the above conditions, are the Ernst equations for an
Einstein-Maxwell-Weyl field. A more standard form of the latter equations is given by a dual 
system, which is obtained here by rewriting the system (\ref{eq:sys1})-(\ref{eq:sys3}) in terms of
the above reality conditions. These conditions imply the 
following constraints on the variables $(a_i,b_i)$:  	
\begin{equation} 
a_1=b_1+\frac{v-u}{f_1+f_1^{\star}+f_2 f_2^{\star}}\,,\quad 
a_2=b_2+\frac{(v-u)f_2^{\star}}{f_1+f_1^{\star}+f_2 f_2^{\star}}\,. \label{eq:Gs}
\end{equation}
By virtue of (\ref{eq:Gs}), one readily notices that the system of equations 
(\ref{eq:sys1})-(\ref{eq:sys3}), decouples for the variables $(f_1,f_2)$, leading
to a second order system. In order to make contact with the notation employed in 
general relativity, we relabel the main complex variables as $(f_1,f_2)\equiv({E},\Phi)$. 
Using equation (\ref{eq:Gs}) and adopting a coordinate free setting,
the relevant second order system of PDEs is given by equations (\ref{eq:sys3}), and reads 
\begin{eqnarray}
({E}+{E}^\star + \Phi \Phi^\star) 
\big(\dif (\varrho \ast \dif {E}) - {\rm i}\, \boldsymbol{\beta} \wedge \dif {E}\big)&=&  
2 \varrho (\dif {E} + \Phi^\star \dif \Phi)\wedge \ast \dif {E},
\label{eq:ernst1}\\
({E}+{E}^\star + \Phi \Phi^\star) \big(\dif (\varrho \ast \dif \Phi) - 
{\rm i}\, \boldsymbol{\beta} \wedge \dif \Phi\big)&=& 
2 \varrho (\dif {E} + \Phi^\star \dif \Phi)\wedge \ast 
\dif \Phi\, .\label{eq:ernst2}
\end{eqnarray} 
Here, ${\rm{i}}=\sqrt{-1}$ and the one-form  $\boldsymbol{\beta}$ is given by 
\begin{equation}
\boldsymbol{\beta}=\nu(u)\dif u + \mu(v) \dif v\, ,
\end{equation}
where the real parameter functions the $\nu,\mu$, are introduced by recasting
the parameter constraints (\ref{eq:nm}) in the form
\begin{equation} \textstyle
n= -\frac{1}{2} + {\rm{i}} \nu(u)\,,\quad m= -\frac{1}{2} + {\rm{i}} \mu(v)\,.
\label{eq:parameters}
\end{equation} 

In the context of general relativity, equations (\ref{eq:ernst1}), (\ref{eq:ernst2}) 
are known as the Ernst equations for an Einstein-Maxwell-Weyl field \cite{Alex2}.
It should be clear from the preceding discussion that, all solutions of the 
Ernst equations are embedded into the solution space of the Euler-Lagrange 
equations (\ref{eq:ELeqs}), while the converse does {\em not} hold in general. 
Therefore, equations (\ref{eq:ELeqs}), or equivalently the auxiliary 
system (\ref{eq:sys1})-(\ref{eq:sys2}), represent a generalization of the Ernst
equations (\ref{eq:ernst1}), (\ref{eq:ernst2}).

At this point, it is worth noticing that after applying the conditions 
(\ref{eq:reality}), (\ref{eq:parameters}) and their consequences 
to the linear representation (\ref{eq:LA}), the latter 
reduces to the one introduced in \cite{Alex2,Alex3,Alex4}.

\section{Discussion}
Apart from the Ernst equations, which possess a prominent position among the 
subsystems included into the solution space of equations (\ref{eq:ELeqs}), 
there are also various interesting subsystems which are tied in with the latter
set of equations. Thereof, we point out that equations (\ref{eq:ELeqs}), 
incorporate also the PDE which encodes the KdV hierarchy. This is accomplished 
by considering the case, $f^1=f^2\equiv f$, which leads to a scalar fourth order
PDE for the variable $f$. The latter equation is the Euler-Lagrange equations 
for the Lagrangian density
\begin{equation} 
{\mathcal L} = (u-v) \frac{(f_{,uv})^{\,2}}{f_{,u}f_{,v}} + 
\frac{1}{u-v} \left( m^2 \frac{f_{,u}}{f_{,v}} + 
n^2 \frac{f_{,v}}{f_{,u}} \right)\, . \label{eq:KdV}
\end{equation}
In \cite{NHJ}, \cite{TTX} it was shown that in the case where $n$ and $m$ are 
constants, the corresponding PDE for $f$, admits similarity solutions built from 
solutions of the full Painlev\'e VI (PVI) equation, i.e. with all parameters free. 
Alternatively, full PVI can be shown to arise naturally as symmetry reduction 
from the anti-self-dual Yang-Mills equations \cite{MasonWoodhousebook}, 
or from the Ernst equation for an Einstein-Weyl field \cite{Schief2}. Taking
into account the richness of equations (\ref{eq:ELeqs}), 
it is even more interesting to consider their similarity 
reductions. Such a reduction leads to a six-parameter, second order, coupled 
system of ordinary differential equations \cite{TN}, which we expect to 
be associated with the hierarchies of the PVI equation \cite{NW}. 
  
\subsection*{Acknowledgments}
The initial stage of the work of A.T. was supported by the University of 
Patras research project C. Carath\'eodory, 2001/2785 and successively and mainly
by the European postdoctoral fellowship Marie Curie, contract No 
HPMF-CT-2002-01639. The authors thank the referee for helpful suggestions.


\end{document}